# Effect of high temperature annealing (T > 1650°C) on the morphological and electrical properties of p-type implanted 4H-SiC layers

M. Spera [1,2,3], D. Corso [1], S. Di Franco [1], G. Greco [1], A. Severino [4], P. Fiorenza [1], F. Giannazzo [1], F. Roccaforte [1]*

[1] CNR-IMM, Strada VIII n. 5 Zona - Industriale - 95121 Catania, Italy
[2] Department of Physics and Astronomy, University of Catania, via Santa Sofia 64, 95123, Catania, Italy
[3] Department of Physics and Chemistry, University of Palermo, via Archirafi 36, 90123, Palermo, Italy
[4] STMicroelectronics, Stradale Primosole, 50, 95121 Catania, Italy

* e-mail: fabrizio.roccaforte@imm.cnr.it

**Abstract**

This work reports on the effect of high temperature annealing on the electrical properties of p-type implanted 4H-SiC. Ion implantations of Aluminium (Al) at different energies (30 – 200 keV) were carried out to achieve 300 nm thick acceptor box profiles with a concentration of about $10^{20}$ at/cm$^3$. The implanted samples were annealed at high temperatures (1675-1825 °C). Morphological analyses of the annealed samples revealed only a slight increase of the surface roughness RMS up to 1775°C, while this increase becomes more significant at 1825°C (RMS=1.2nm). Room temperature Hall measurements resulted in a hole concentration in the range 0.65-1.34×$10^{18}$/cm$^3$ and mobility values in the order of 21-27 cm$^2$V$^{-1}$s$^{-1}$. The temperature dependent electrical measurements allowed to estimate an activation energy of the Al-implanted specie of about 110 meV (for the post-implantation annealing at 1675°C) and a fraction of active p-type Al-dopant ranging between 39% and 56%. The results give useful indications for the fabrication of 4H-SiC JBS and MOSFETs.

**Introduction**

Silicon carbide (4H-SiC) is by now considered as the best semiconductor for energy efficient modern power electronics [1]. In fact, the outstanding physical and electronic properties of 4H-SiC lead to enormous benefits in high-temperature, high-power and high-frequency applications, allowing to significantly reduce the power losses with respect to the traditional silicon devices [2,3].

Because of the extremely low diffusion coefficients of the dopant species in SiC even at high temperatures, ion implantation is practically the unique choice to achieve selective area doping of 4H-SiC during devices fabrication [1,4]. As a matter of fact, today, all the commercially available 4H-SiC power devices, i.e., Schottky diodes, Junction Barrier Schottky (JBS) diodes, metal oxide semiconductor field effect transistors (MOSFETs), employ ion-implantation for selective area doping or for creating resistive edge termination structures [1].

The main dopant species for SiC are Nitrogen (N) and Phosphorous (P) for n-type doping, and Aluminum (Al) for p-type doping. High post-implantation annealing temperatures (> 1500°C) are typically required to bring these species in substitutional positions and achieve their electrical activation [5,6,7]. In particular, selectively doped p-type regions are key parts of both JBS and MOSFETs and the control of their electrical properties has a significant impact on several devices parameters (e.g., Ohmic contacts formation, device on-resistance, threshold voltage and channel mobility, etc.).

Hence, understanding the dependence of the properties of p-type implanted layers on the activation annealing temperature is very important for device manufacturers to set the right process for the optimal device characteristics. In this context, although several investigations reported on the properties of Al-implanted 4H-SiC layers [8,9,10], the large variety of experimental conditions and the evolution of the annealing procedures always make this topic open to scientific discussions.

Hall-effect measurements are often used to study the electrical properties of p-type 4H-SiC layers, in order to determine key parameters like the holes concentration and mobility [11]. A critical issue of this methodology is the choice of the Hall scattering factor $r_H$ for SiC [12,13]. In fact, the difficulty to extract the mobility and the free hole concentration from Hall measurements is related to the correct knowledge of $r_H$. However, many authors often interpret the experimental Hall results on p-type 4H-SiC assuming $r_H=1$, which in turn leads to an overestimation of the doping level in the material [9,14]. Only few works specifically reported experimental calculations of the Hall scattering factor $r_H$ for SiC [15,16], whose findings should be considered for a correct analysis of the currently available data.

This paper reports on the morphological and electrical properties of p-type implanted 4H-SiC annealed at three different temperatures for electrical activation. The activation of the p-type Al-dopant has been investigated by means of Van der Pauw and Hall effect measurements, taking into account the concentration dependent Hall scattering factor $r_H$ [15]. From the temperature dependent Hall measurements, the activation energy and the active fraction of p-type Al-dopant have been extracted as a function of the annealing temperature. The results have been discussed considering the possible implications in practical devices fabrication.

**Experimental Details**

N-type doped 4H-SiC (0001) epitaxial layers grown onto heavily doped substrates with a nominal doping concentration of $N_{D-epi}=1\times10^{16}$ at/cm$^3$ were used in this study.
The samples were implanted at 500°C with Al-ions at different energies (30-200 keV) and doses of $3\times10^{14} - 1\times10^{15}$ at/cm$^2$ in order to obtain an almost flat (box-like) profile with a thickness $t_{imp}$ = 300 nm and a concentration of $1\times10^{20}$ at/cm$^3$. After implantation, the samples were annealed at three different temperatures (1675°C for 30 minutes, 1775°C and 1825°C for 15 minutes). During annealing, the sample surface was protected by a graphite capping layer, formed by an appropriate thermal backing of photoresist [17]. The same cooling rate of 25 °C/min has been used for the three samples after the high temperature annealing treatment. In fact, keeping a slow cooling rate is required in order to avoid the formation of a large amount of carbon vacancies ($V_C$) during elevate heat treatments in 4H-SiC [18].

The surface morphology was monitored by means of Atomic Force Microscopy (AFM), using a PSIA XE-150 microscope. In order to determine the electrical properties of the p-type implanted layers, Hall measurements were carried out in the temperature range 300-500 K, using a MMR Hall Effect Measurement System H-50 equipment. The Hall structures were fabricated by defining isolated p-type implanted square structures and contacting their four corners by Ti (70 nm)/Al (200 nm) Ohmic contacts annealed at 950°C.

**Results and Discussion**

Fig 1 shows the AFM images of p-type implanted 4H-SiC samples, subjected to tree different annealing temperatures. As can be seen, the root mean square (RMS) roughness of the sample annealed at 1675°C is 0.17 nm, i.e. which is similar to the typical values measured in the unannealed layers. However, the RMS increases with increasing the annealing temperature (e.g., RMS=0.24 nm at 1775°C) until a notable degradation of the surface morphology is observed after annealing at 1825°C (RMS=1.2 nm). In this latter case, the sample surface is characterized by the presence of holes that reach a depth of about 8-10 nm and a diameter of about 200 nm. A similar surface morphology degradation has been previously attributed to the damage occurring in the graphite capping layer during annealing at such high temperatures [19]. Hence, a further optimization our process could be required for its implementation in a real device fabrication flow.

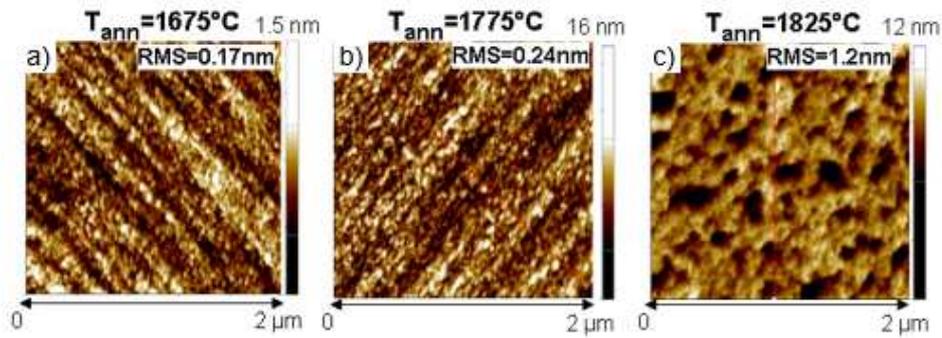

Fig. 1. AFM images of p-type implanted 4H-SiC layers after activation annealing at different temperatures $T_{ann}$: (a) 1675°C, (b) 1775°C and (c) 1825°C.

For the electrical characterization of the implanted layers, Van der Pauw and Hall-effect measurements have been performed on the samples annealed at different temperatures. First, the current-voltage (I-V) curves (Fig. 2a) have been acquired on Van der Pauw structures fabricated on the three implanted samples (inset of Fig. 2a). The value of resistance (R) extracted from the slope of the I-V curves has been used to calculate the sheet resistance of the implanted layer ($R_{SH}$), according to the equation [13]

$$R_{SH} = \frac{\pi}{\ln 2} R \qquad (1)$$

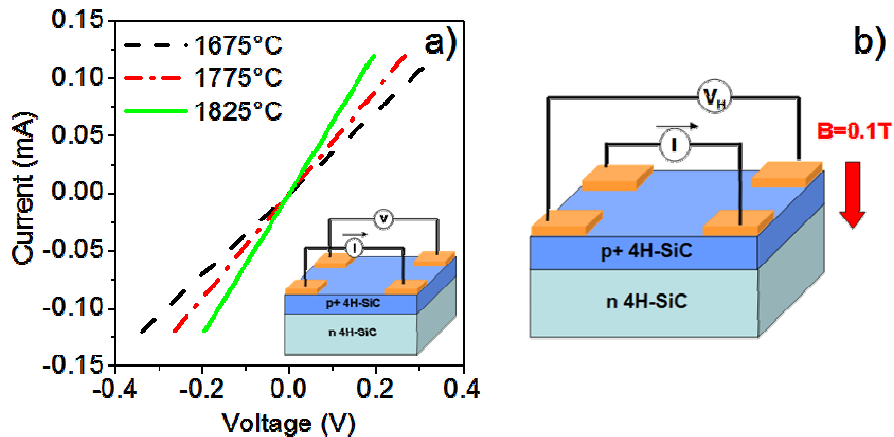

**Fig. 2**: (a) I-V curves acquired on Van der Pauw structures (inset) fabricated on p-type implanted 4H-SiC samples annealed at three different temperatures: 1675°C, 1775°C and 1825°C. (b) Schematic of the Hall effect measurement.

As can be seen in Fig. 2a, the slope of the I-V curves increases with increasing the annealing temperature, thus indicating a decrease of the value of $R_{SH}$. In fact, from Eq. (1) the following $R_{SH}$ values have been extracted: 12.0 kΩ/sq (1675°C), 9.75 kΩ/sq (1775°C) and 7.47 kΩ/sq (1825°C). By assuming a uniform doping concentration in the entire implanted layer thickness $t_{imp}$, it was possible to quantify the decrease of the resistivity of the p-type 4H-SiC upon annealing, i.e., from ρ = 0.36 Ωcm (1675°C) down to ρ = 0.22 Ωcm (1825°C). The obtained data are reported in Table 1.

The dependence of the resistivity on the annealing temperature of the implanted layer provides useful information on the mechanism of activation of the implanted dopant. In particular, an Arrhenius plot of the sheet resistance $R_{SH}$ can allow to estimate an activation energy for the electrical activation of the dopant, i.e., the energy needed to Al atoms to contribute as acceptors to the p-type doping of the 4H-SiC implanted layer. Fig. 3 reports an Arrhenius plot of the sheet resistance of the Al-implanted 4H-SiC layers as a function of the annealing temperature. The linear fit of the data results into an activation energy of about 1eV. In the same graph, other literature data referring to 4H-SiC layers implanted with a similar Al content ($1-3\times 10^{20}$cm$^{-3}$) are reported [20]. As can be seen, the linear fits of the data exhibit almost the same activation energy, thus confirming that the mechanism of electrical activation of Al for high concentrations (i.e., in the order of $10^{20}$cm$^{-3}$) is the same in the two cases. *Giannazzo et al.* [21] reported a value of the activation energy for Al electrical activation of 6.3eV . However, this strong discrepancy is fully justified by the fact that room temperature Al-implantation and annealing temperatures from 1550 to 1650 °C have been considered in Ref. [21] to extract the activation energy.

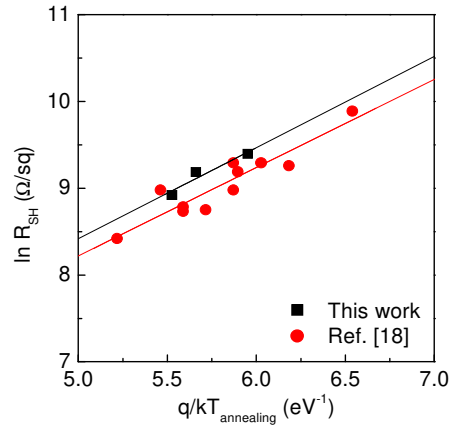

**Fig. 3**: Arrhenius plot of the sheet resistance of the Al-implanted 4H-SiC samples as a function of the annealing temperature, giving an activation energy of 1.eV. For comparison, a similar Arrhenius plot for the data reported in Ref. [20] is also reported.

Thereafter, a magnetic field of intensity B = 0.1 T, perpendicular to the sample surface, has been applied to perform Hall-effect measurements, as can be seen from the schematic reported in Fig. 2b. By fixing the value of the current *I*, it is possible to measure a Hall voltage $V_H$ and to determine the density *p* and the sign of the carriers from the following equation:

$$p = \frac{IB}{V_H q t_{imp}} \qquad (2)$$

where q is the elementary charge and the term $R_H = \dfrac{t_{imp} V_H}{BI}$ is the so called Hall coefficient.

Then, combining the values of the sheet resistance $R_{SH}$ and of the carrier density $p$, it is possible to determine the Hall mobility $\mu$ as :

$$\mu = \frac{1}{qpR_{SH}t_{imp}} \qquad (3)$$

Using Eq. (2) and Eq. (3) the following values have been extracted: $p=1.38\times10^{18}$ cm$^{-3}$ and $\mu=13$ cm$^2$/Vs (1675°C), $p=2.29\times10^{18}$ cm$^{-3}$ and $\mu=10.7$ cm$^2$/Vs (1775°C), $p=2.58\times10^{18}$ cm$^{-3}$ and $\mu=10.9$ cm$^2$/Vs (1825°C). These values of p and µ at room temperature have been evaluated, in a first approximation, assuming a Hall scattering factor $r_H = 1$. The corrected values taking into account an appropriate $r_H$ for our experimental conditions will be reported later on in this paper.

In order to determine the electrical activation and the compensation ratio of the p-type implanted dopant under different annealing conditions, temperature dependent Hall measurements have been carried out.
In particular, in a p-type semiconductor the hole concentration $p$ as a function of the temperature can be described by the neutrality equation [13]:

$$\frac{p(p+N_D)-n_i^2}{N_A-N_D-p+n_i^2/p} = \frac{N_V}{g}\exp\left(-\frac{E_A}{kT}\right) \qquad (4)$$

where $N_A$ is the acceptor concentration, $N_D$ is the compensating donor concentration, $N_V$ is the density of states in the valence band, $E_A$ is the activation energy (i.e., the energy level of the acceptors above the valence band) and $g$ is the degeneracy factor of the ground level of the Al-acceptor fixed to 4 [15].
Considering that for wide band gap semiconductors the intrinsic carrier concentration $n_i$ is extremely low, the expression of the hole concentration in Eq. (4) can be approximated by [10]:

$$p \approx \frac{1}{2}\left[-N_D - x + \sqrt{(N_D-x)^2 + 4N_A x}\right] \qquad (5)$$

where

$$x = \frac{N_V}{g}\exp\left(-\frac{E_A}{kT}\right) \qquad (6)$$

As specified in the introduction, one of the methodology issues encountered in the correct determination of the hole concentration $p$ is to take into account the Hall scattering factor $r_H$. In fact, the hole concentration $p$ determined by Hall measurements is related to the Hall coefficient $R_H$ by the expression [13]:

$$p = \frac{r_H}{qR_H} \qquad (7)$$

Hence, by simply assuming $r_H = 1$ typically leads to an overestimation of the carrier density in SiC [11]. To overcome this problem, firstly *Pensl et al.* [15] proposed to derive the Hall scattering factor $r_H$ for p-type epitaxial SiC from the comparison between the temperature dependent experimental and theoretical values of $p$:

$$r_H = \frac{p_{theor}(T)}{p_{exp}(T)} \tag{8}$$

where $p_{exp}(T)$ is the experimental hole concentration obtained assuming $r_H=1$ and $p_{theor}(T)$ is the theoretical value calculated from the neutrality Eq. (5) using values of $N_A$ and $N_D$ obtained by independent chemical and electrical analyses [15].

Then, in order to estimate the correct value of carrier concentration from the Hall measurement, the Hall scattering values should be considered. As first approach, the temperature dependent Hall scattering factor $r_H$ reported by *Pensl et al.* [15] was considered to analyze our results, thus resulting into values of $N_A$ larger than the total Al-implanted concentration when fitting the experimental data with the neutrality equation. However, such an overestimation could be related to the fact that the values of $r_H$ reported in Ref. [15] were determined for Al-doping levels of about $1\times10^{18}$ cm$^{-3}$, i.e., about two orders of magnitude smaller that the Al-implanted concentration of our samples ($1\times10^{20}$ cm$^{-3}$).

More recently, *Asada et al.* [16] derived the Hall scattering factor $r_H$ for a wider range of Al-concentration in 4H-SiC epilayers.. Hence, we considered these results [16] to correct our experimental data for all the measurement temperatures. Namely, the experimental values of the hole concentration $p_{exp.}(T)$ determined by temperature dependent Hall measurements were corrected using the expression:

$$p_{corr.}(T) = p_{exp.}(T) \cdot r_H(T) \tag{9}$$

Fig. 4 shows, as an example, the values of the hole concentration determined from Hall-effect measurements on the sample annealed at 1675°C assuming $r_H=1$ (black squares) and after correction (red circles). As can be seen in Fig. 4, a significant decrease of the hole concentration occurs after correction, since in the temperature range of our experimental measurements (300-500K) the values of $r_H$ decrease from 0.5 to 0.25 [16]

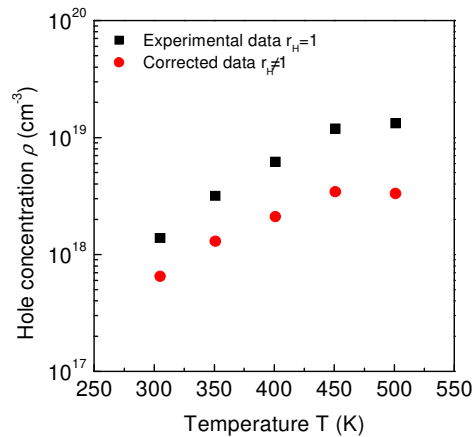

**Fig. 4**: Experimental data ($r_H=1$) of the hole concentration measured on the Al- implanted 4H-SiC sample annealed at 1675°C (black squares) and data corrected ($r_H\neq1$) using the $r_H$ factor reported in Ref. [16] (red circles).

Similarly, the experimental Hall mobility values determined from Eq. (3) as a function of the measurement temperature have been corrected considering the Hall scattering factor. Then, the same correction has been applied also for the other samples annealed at 1775°C and 1825°C.

Table 1 summarizes all the physical parameters calculated by Van der Pauw and Hall effect measurements for the three samples at room temperature.

| Annealing Temperature | $R_{SH}$ (Ω/sq) | ρ (Ωcm) | p (cm$^{-3}$) | μ (cm$^2$V$^{-1}$s$^{-1}$) |
|---|---|---|---|---|
| 1675°C | 12000 | 0.36 | 6.49×10$^{17}$ | 26.7 |
| 1775°C | 9750 | 0.29 | 9.68×10$^{17}$ | 22.1 |
| 1825°C | 7470 | 0.22 | 1.34×10$^{18}$ | 20.8 |

**Table 1**: Summary of the parameters extracted by Van der Pauw and Hall.effect measurements at room temperature on p-type implanted 4H-SiC samples, annealed at three different temperatures $T_{ann}$: 1675°C, 1775°C and 1825°C.

The Hall mobility values are in the range 21-27 cm$^2$/Vs for these annealing temperatures, that well agree with the results reported by *Rambach et al.* [9] for similar temperature of implantation.
In order to determine the values of $N_A$ and $N_D$, the application of the neutrality equation (Eq. (5)) requires the knowledge of the ionization energy of the Al acceptors.
An estimation of $E_A$ can be made considering the temperature dependence of the sheet resistance $R_{SH}$ (T). In fact, the temperature dependent sheet resistance $R_{SH}(T)$ of an implanted layer arises from both the temperature dependence of mobility μ(T) and carrier concentration p(T), i.e.,:

$$R_{SH}(T) = \frac{1}{q\mu(T)p(T)t_{imp}} \quad (10)$$

where $t_{imp}$ is the thickness of the implanted layer.
From the neutrality equation (see Eqs. (5) and (6)), it can be deduced that the temperature dependence of the carrier concentration is dominated by an exponential of the ionization energy of the Al-dopant $E_A$:

$$p(T) \propto \exp\left(-\frac{E_A}{kT}\right) \quad (11)$$

On the other hand, the carrier mobility shows a much weaker dependence on the measurement temperature [17]. Hence, it is possible to approximate the temperature dependence of the sheet resistance as:

$$R_{SH}(T) \propto \frac{1}{\exp(-E_A/kT)} \quad (12)$$

Under this approximation, the value of the ionization energy $E_A$ can be estimated from a linear fit of the plot of ln($R_{SH}$) as a function of $q/kT$. As an example, Fig. 5 shows this plot for the sample annealed at 1675°C. From a linear fit of the data, an activation energy $E_A$=110 meV could be determined.

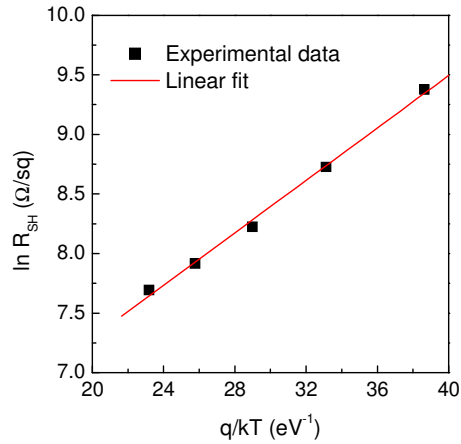

**Fig. 5:** Linear fit of the logarithmic values of $R_{SH}$ as a function of $q/kT$ for the sample annealed at 1675°C. From the linear fit, an activation energy $E_A$=110 meV could be determined.

Following this method, it was possible to determine the values of $E_A$ for the other samples annealed at higher temperatures, i.e., $E_A$=105meV and $E_A$=99meV for the sample annealed at 1775°C and 1825°C, respectively. This decrease of the ionization energy of the Al-dopant with the annealing temperature will be discussed later.

Fig. 6a-c reports the temperature dependence of the experimental values of the hole concentration for the three samples annealed at 1675°C, 1775°C and 1825°C. The continuous lines represent the fit of the experimental data with the Eq. (5), assuming the values of $E_A$ estimated by the temperature dependence of $R_{SH}$ (as the example in Fig. 5) and considering the acceptor concentration $N_A$ and the compensating donor concentration $N_D$ as free parameters. Table 2 reports the acceptor and donor concentration values determined from the fits of the experimental data, together with the value of the ionization energy $E_A$. From Table 2, it can be observed that, as the annealing temperature increases, the acceptor concentration increases, thus being accompanied by a slight decrease of the compensating donors concentration. The obtained values result in an increase of the Al activation from 39% to 56%, with increasing annealing temperature from 1675°C to 1825°C, consistently with the decrease of the sample resistivity. Similar activation values have been obtained from *Negoro et al.* [5] considering an annealing temperature of 1800°C and similar values of Al implanted concentration and temperature of implantation. However, *Nipoti et al.* [22], reported a higher activation values, of 69%, using a microwave system which reaches an annealing temperature of 2100°C. In our case, the estimated activation percentage can be justified by the lower annealing temperature. In fact, for a value of 1750°C, an activation of 35% has been reported from *Saks et al.* [23]. Moreover, it has been observed that an improvement of the Al activation (up to 100%) can be obtained by increasing the implantation temperature up to 1000°C [9].

As could be deduced from our analysis, an Al activation of 56% could be reached with an increase of the annealing temperature up to 1825 °C. At his temperature, however, a degradation of the surface morphology has been observed by AFM (Fig. 1c), which in turn may prevent the practical use of such doping conditions in MOSFETs technology. However, in other cases, such as in the case of JBS or bipolar devices, the need of a high acceptor concentration is more important to minimize the specific contact resistance of Ohmic contacts on the implanted regions. In this case, considering that the current transport in metal/p-type 4H-SiC interfaces is typically ruled by Thermionic Field Emission mechanism [24] and assuming a barrier height of 0.56 eV (for a Ti/Al-based contact) [25], the increase of the acceptor concentration from $3.9 \times 10^{19}$ cm$^{-3}$ to $5.6 \times 10^{19}$ cm$^{-3}$ should result in an improvement of the specific contact resistance of about the 82 % (i.e. from $1.8 \times 10^{-4}$ Ωcm$^2$ to $3.2 \times 10^{-5}$ Ωcm$^2$).

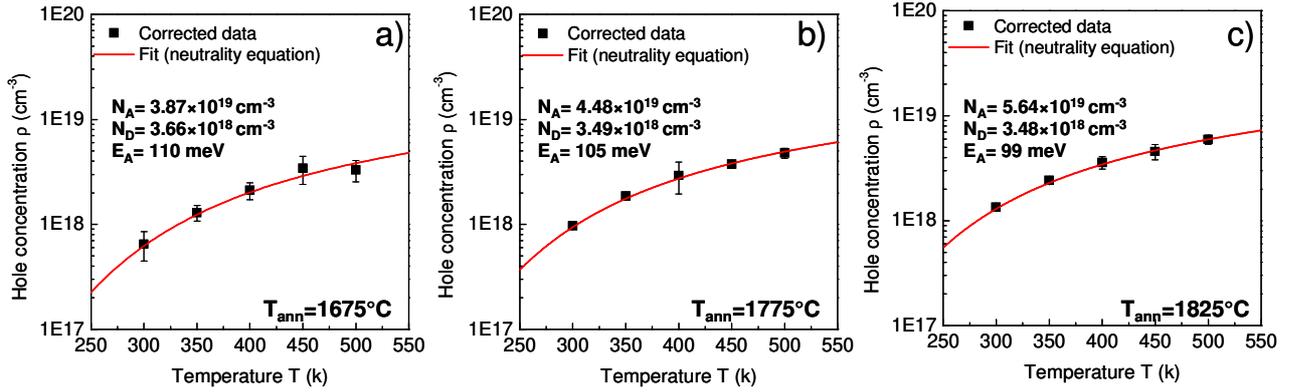

**Fig. 6**: Temperature dependence of the hole concentration for the Al-implanted 4H-SiC samples annealed at $T_{ann}$=1675°C (a)), $T_{ann}$=1775°C (b)) and $T_{ann}$=1825°C (c). The continuous line represents the fits of the experimental data with the neutrality equation.

| Annealing Temperature | $E_A$ (meV) | $N_A$ (cm$^{-3}$) | $N_D$ (cm$^{-3}$) | Al activation | Compensation |
|---|---|---|---|---|---|
| 1675°C | 110 | 3.87×10$^{19}$ | 3.66×10$^{18}$ | 39 % | 9.4 % |
| 1775°C | 105 | 4.84×10$^{19}$ | 3.49×10$^{18}$ | 48 % | 7.2 % |
| 1825°C | 99 | 5.64×10$^{19}$ | 3.48×10$^{18}$ | 56 % | 6.2 % |

**Table 2**: Values of activation energy $E_A$, acceptor concentration $N_A$, compensating donor concentration $N_D$, Al activation and compensation ratio for the Al-implanted 4H-SiC samples annealed at 1675°C, 1775°C and 1825°C.

Finally, as mentioned before and reported in Table 2, the values of the ionization energy $E_A$ decreases with increasing the acceptor concentration $N_A$. In fact, it is known that $E_A$ is inversely proportional to the average distance between Al atoms, that in turn decreases with the increase of the acceptor density $N_A$ [11].
In general, this dependence of the ionization energy on the acceptor concentration is described by the empirical expression [26]:

$$E_A = E_0 - \alpha \cdot N_A^{1/3} \tag{13}$$

where $E_0$ is the ionization energy for an isolated impurity centre, and α is a constant.
Then, Fig. 7 reports in a plot of $E_A$ as a function of $N_A$ our experimental data together with other literature works [8] [12] on Al-implanted 4H-SiC layers. This collection of experimental data has been fitted with the theoretical behaviour of $E_A$ as function of $N_A$, expressed by the Eq. (13), giving a ionization energy $E_0$=216 meV and a coefficient α = 3×10$^{-5}$ meV cm$^{-1}$.
Noteworthy, *Frazzetto et al.* [17] determined an activation energy 144 meV and a doping level of 2×10$^{-19}$ cm$^{-3}$ by modelling the temperature dependence of the specific contact resistance of Ohmic contacts on p-type implanted 4H-SiC layers using transmission line model (TLM) structures. This data, reported in Fig. 7, is also well described by the experimental fit obtained using Eq. (14). This results suggests that an appropriate analysis of TLM data can also provide useful information on the electrical activation of Al dopant in 4H-SiC..

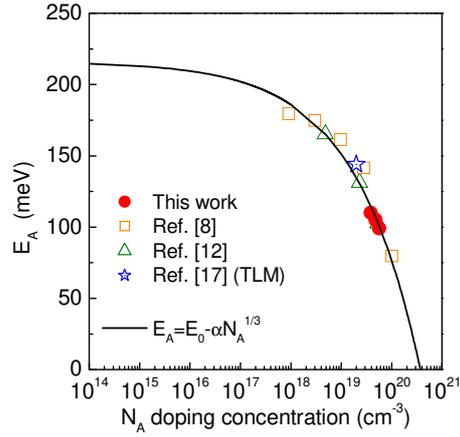

**Fig. 7:** Ionization energy of the acceptors ($E_A$) in 4H-SiC as a function of the acceptor concentration ($N_A$) determined by Hall measurements. The experimental data of this work (red circles) are reported together with data from Refs. [8] [12] [17]. The solid line corresponds to a fit with the empirical expression $E_A = E_0 - \alpha N_A^{1/3}$ with $E_0 = 216$ meV and $\alpha = 3 \times 10^{-5}$ meV cm$^{-1}$.

**Conclusion**

In summary, in this paper the morphological and electrical behaviour of Al-implanted 4H-SiC annealed at different temperatures was investigated.
The morphological analyses of the annealed samples indicated that the surface roughness is well preserved upon annealing at 1775°C, while an increase of the RMS (1.2nm) occurs at 1825°C, due to a degradation of the graphite capping layer during annealing at such high temperatures.
Van der Pauw and Hall-effect measurements have been carried out as a function of temperature, taking into account the previously reported Hall scattering factors values. These analyses allowed to determine an active Al-doping concentration of $3.87 \times 10^{19}$ cm$^{-3}$ after annealing at 1675°C, corresponding to an activation of 39%. The electrical activation can increase up to 56% with increasing the annealing temperature up to 1825°C. This latter can be useful for JBS technology to improve the quality of Ohmic contacts. However, further optimization of the capping layer process is required to prevent the RMS increase of Al-implanted 4H-SiC regions upon annealing at such high temperature. Finally, the dependence of the ionization energy $E_A$ on the acceptor concentration $N_A$ has been discussed comparing our results with other literature Hall data and TLM measurements.

**Acknowledgements**

The authors would like to acknowledge the colleagues of STMicroelectronics for their valuable technical support during sample implantation and annealing. STMicroelectronics is also acknowledged for financial support under research contract # 2017-4860 "Study and realization of active components for power devices in SiC". Part of this research activity has been carried out in the framework of the European ECSEL JU project REACTION (grant agreement n. 783158).

# References

[1] T. Kimoto, J. Cooper, *Fundamentals of Silicon Carbide Technology: Growth, Characterization, Devices and Applications*, John Wiley & Sons, Singapore Pte. Ltd., 2014.
[2] F. Roccaforte, P. Fiorenza, G. Greco, R. Lo Nigro, F. Giannazzo, A. Patti, M. Saggio, *Challenges for energy efficient wide band gap semiconductor power devices*, Phys. Status Solidi A 211, (2014) 2063-2071.
[3] F. Roccaforte, P. Fiorenza, G. Greco, R. Lo Nigro, F. Giannazzo, F. Iucolano, M. Saggio, *Emerging trends in wide band gap semiconductors (SiC and GaN) technology for power devices*, Microelectronic Engineering 187-188, (2018) 66-77.
[4] J. Wong-Leung, M.S. Janson, A. Kuznetsov, B.G. Svensson, M.K. Linnarsson, A. Hallén, C. Jagadish, D.J.H. Cockayne, *Ion Implantation in 4H-SiC*, Nucl. Istr. Meth. B 266, (2008) 1367-1372.
[5] Y. Negoro, T. Kimoto, H. Matsunami, F. Schmid, G. Pensl, *Electrical activation of high-concentration aluminum implanted in 4H-SiC*, J. Appl. Phys. 96, (2004) 4916-4922.
[6] H. Fujihara, J. Suda, T. Kimoto, *Electrical properties of n- and p-type 4H-SiC formed by ion implantation into high-purity semi-insulating substrates*, Jpn. J. Appl. Phys. 56, (2017) 070306.
[7] R. Nipoti, H.M. Ayedh, B.G. Svensson, *Defects related to electrical doping of 4H-SiC by ion implantation*, Mater. Sci. Semicond. Proc. 78, (2018) 13-21.
[8] J. Pernot, S. Contreras, J. Camassel, "*Electrical transport properties of aluminium-implanted 4H-SiC*", J. Appl. Phys. 98 (2005) 023706.
[9] M. Rambach, A.J. Bauer, H. Ryssel, "*Electrical and topographical characterization of aluminium implanted layers in 4H silicon carbide*", phys. stat. sol. (b) 245 (2008) 1315-1326.
[10] M. Rambach, A.J. Bauer, H. Ryssel "*High Temperature Implantation of Aluminum in 4H Silicon Carbide*" Materials Science Forum Vols. 556-557 (2007) pp 587-590
[11] S. Contreras, L. Konczewicz, R. Arvinte H. Peyre1, T. Chassagne M. Zielinski, and S. Juillaguet "*Electrical transport properties of p-type 4H-SiC*" Phys. Status Solidi A 214, No. 4, 1600679 (2017)
[12] A. Parisini, R. Nipoti, "*Analysis of the hole transport through valence band states in heavy Al doped 4H-SiC by ion implantation*", J. Appl. Phys. 114 (2013) 243703.
[13] D. K. Schroder, T. T. Braggins, and H. M. Hobgood, "*The doping concentration of indium-doped silicon measured by Hall, C-V and junction breakdopwn techniques*" J. Appl. Phys. 49, 5256 (1978)
[14] A. Koizumi, J. Suda, T. Kimoto, *Temperature and doping dependence of electrical properties of Al-doped 4H-SiC epitaxial layers*, J. Appl. Phys. 106 (2009) 013716.
[15] G. Pensl, F. Schmid, F. Ciobanu, M. Laube, S. A. Reshanov, N. Schulze, K. Semmelroth, H. Nagasawa, A. Schöner, G. Wagner "*Electrical and optical characterization of SiC*" Mater. Sci.Forum Vols. 433-436 (2003) pp 365-370
[16] S. Asada, T. Okuda, T. Kimoto and J. Suda "*Hall scattering factors in p-type 4H-SiC with various doping concentrations*" Appl. Phys. Express 9 (2016) 041301
[17] A. Frazzetto, F. Giannazzo, R. Lo Nigro, V. Raineri, F. Roccaforte, *Structural and transport properties in alloyed Ti/Al Ohmic contacts formed on p-type Al-implanted 4H-SiC annealed at high temperature*, J. Phys. D: Appl. Phys. 44, (2011) 255302.
[18] H. M. Ayedh, R. Nipoti, A Hallen and B.G. Svensson, *Controlling the carbon vacancy concentration in 4H-SiC subjected to high temperature treatment*, Mater. Sci. Forum, vol 858 pp 414-417, (2016).
[19] K.V. Vassilevski, N.G. Wright, A.B. Horsfall, A.G. O'Neill, M.J. Uren, K.P. Hilton, A.G. Masterton, A.J. Hydes and C.M. Johnson, *Protection of selectively implanted and patterned silicon carbide surfaces with graphite capping layer during post-implantation annealing*, Semicond. Sci. Technol 20, 271 (2005).

[20] R. Nipoti, A. Carnera, G. Alfieri, L. Kranz *About the Electrical Activation of 1×1020 cm-3 Ion Implanted Al in 4H-SiC at Annealing Temperatures in the Range 1500 - 1950°C*, Mater. Sci. Forum, Vol. 924, pp 333-338 (2018)

[21] F. Giannazzo, M. Rambach, D. Salinas, F. Roccaforte, and V. Raineri, *Electrical characterization of Al Implanted 4H-SiC Layers by Four Point Probe and Scanning Capacitance Microscopy*, Mater. Sci. Forum 615-617 (2009) 457-460.

[22] R. Nipoti, A. Nath, M. V. Rao, A. Hallen, A. Carnera, Y. L. Tian, *Microwave annealing of very high dose aluminum implanted 4H-SiC*, Applied Physics Express 4 (2011) 111301

[23] N. S. Saks, A. V. Survov, D. C. Capell, *High temperature high dose implantation of aluminum in 4H-SiC*, Appl. Phys. Lett. 84, 5195 (2004)

[24] F. Roccaforte, M. Vivona, G. Greco, R. Lo Nigro, F. Giannazzo, S. Rascunà, M. Saggio, *Metal/Semiconductor contacts to silicon carbide: physics and technology*, Mater. Sci. Forum 924 (2018) 339-344.

[25] M. Vivona, G. Greco, C. Bongiorno, R. Lo Nigro, S. Scalese, F. Roccaforte, *Electrical and structural properties of surfaces and interfaces in Ti/Al/Ni Ohmic contacts to p-type implanted 4H-SiC*, Appl. Surf. Sc. 420, 331–335 (2017)

[26] G. L. Pearson and J. Bardeen, *Electrical Properties of Pure Silicon and Silicon Alloys Containing Boron and Phosphorus,* Phys. Rev. 75, 865 (1949)